\documentclass[epj]{webofc}
\usepackage[utf8]{inputenc}
\usepackage[varg]{txfonts}   
\usepackage{booktabs}
\usepackage{xcolor}
\definecolor{darkred}{rgb}{0.4,0.0,0.0}
\definecolor{darkgreen}{rgb}{0.0,0.4,0.0}
\definecolor{darkblue}{rgb}{0.0,0.0,0.4}
\usepackage[bookmarks,linktocpage,colorlinks,
    linkcolor = darkred,
    urlcolor  = darkblue,
    citecolor = darkgreen]{hyperref}
\usepackage{graphicx}
\usepackage{subfig}
\usepackage{amsfonts}
\usepackage{amssymb}
\usepackage{amsmath}
\usepackage{mathdots}
\usepackage{placeins}
\usepackage{bbold}
\wocname{EPJ Web of Conferences}
\woctitle{Lattice2017}
\let\l\left
\let\r\right
\let\de\partial
\let\mrm\mathrm
\let\mcal\mathcal

\let\mbb\mathbb

\let\dag\dagger

\def\frac#1#2{ {{#1} \over {#2} }}

\def\eg{\hbox{\em e.g. }}
\def\ie{\hbox{\em i.e. }}

\def\beq{\begin{equation}}
\def\eeq{\end{equation}}

\newcommand{\beqa}{\begin{eqnarray}}
\newcommand{\eeqa}{\end{eqnarray}}
\newcommand{\ba}{\begin{array}}
\newcommand{\ea}{\end{array}}
\newcommand{\bmat}{\begin{pmatrix}}
\newcommand{\emat}{\end{pmatrix}}
\newcommand{\bcas}{\begin{cases}}
\newcommand{\ecas}{\end{cases}}
\newcommand{\muh}{{\hat{\mu}}}
\newcommand{\nuh}{{\hat{\nu}}}

\newcommand{\dd}{{\mathrm{d}}}

\newcommand{\tr}{{\mathrm{Tr}}}
\newcommand{\id}{{\mathbb{1}}}

\begin{document}
%
\selectlanguage{english}
\title{%
Simulating lattice field theories on multiple thimbles
}
\author{%
\firstname{Francesco} \lastname{Di
  Renzo}\inst{1,2}\fnsep\thanks{\email{francesco.direnzo@unipr.it}}
}
\institute{%
D.S.M.F.I. Universit\`a di Parma
\and
I.N.F.N. Gruppo collegato di Parma
}
\abstract{%
Simulating thimble regularization of lattice field theory can be tricky when more than one thimble is to be taken into account. A couple of years ago we proposed a solution for this problem. More recently this solution proved to be effective in the case of 0+1 dimensional QCD. A few lessons we can learnt, including the role of symmetries and general hints on algorithmic solutions. 
}
\maketitle
\section{Introduction}\label{sec:intro}

Thimble regularization of lattice field theories was put forward as a
possible solution to the sign
problem~\cite{Cristoforetti:2012su,Fujii:2013sra}. The solution is very elegant
and in principle it is a fundamental one: the domain of
integration of an integral featuring a (possibly widely)
oscillating phase is deformed in such a way that the latter is turned
into a stationary phase. 
There are actually (at least) a couple of caveat attached to the former
statement. First of all, by changing the domain of integration one is
left with a new phase~\cite{Cristoforetti:2014gsa} ({\em aka} the residual phase) which roughly
speaking comes as a
consequence of having changed integration variables. This is not a
really serious problem, or at least for every problem which has been 
worked out till
now in this framework it was shown that this phase can be
effectively taken into account by reweighting. A second matter of
concern is more serious: an integral is in general turned into a sum
of integrals. One speaks of thimble decomposition as the original domain
of integration is turned into the sum of many thimbles. While there is
an argument suggesting that in the thermodynamic limit one single
(dominant) thimble provides the relevant contribution one is
interested in~\cite{Cristoforetti:2012su}, in recent times examples
were provided showing that in finite systems collecting the
contributions of different thimbles can be a tricky 
business~\cite{Alexandru:2015sua,Fujii:2015vha,Tanizaki:2015rda} and this was the main motivation
for the modified approach known as the holomorphic
flow~\cite{Alexandru:2015sua}. \\
The thimble regularization of QCD in $0+1$ dimensions was presented at
the last year Lattice conference~\cite{DiRenzo:2016pwd}. Here we improve on that
work, showing a better way to take into account the three
contributions which are expected in the thimble decomposition of the
problem at hand. Results are better than the previous ones due to two
improvements: first of all, a symmetry argument can reduce 
the number of contributions that we have to sum to solve the theory
(it turns out that there are essentially two distinct contributions); 
in the second place, we make use of a better Monte Carlo
strategy. These are indeed valuable improvements. Together with a 
semiclassical argument pointing out in which
regions of the parameter space it is essential to collect the results
from all the three thimbles, they lead us to the full solution of the
problem. All in all, QCD in $0+1$ dimensions provides a
nice example of how a theory can be simulated on multiple thimbles.

\section{Thimble regularization in a nutshell}\label{sec:thimbles}

\subsection{Thimble decomposition}\label{sec:thimbles_1}
  
A sign problem is in place when one has to solve a theory, \ie 
one has to compute the relevant
functional integrals (we make use of a one-dimensional notation:
results are to be thought of in multiple dimensions)
$$
\frac{1}{Z}\;\int_{-\infty}^{\infty} \mbox{d}x \, O(x) \, e^{-S(x)}
$$
and the action is complex: $S(x) = S_R(x) \, + \, i S_I(x)$. Thimble
regularization is built out of three steps:
\begin{itemize}
  \item We complexify the degrees of freedom: $x \rightarrow z = x + i
    y$ and as a consequence we consider $S(z) = S_R(x,y) \, + \, S_I(x,y)$.
  \item We look for (stationary) critical points satisfying $\partial_z S = 0$.
  \item To each critical point $p_\sigma$ a thimble $\mcal{J}_\sigma$
    is attached and the original function integral is now decomposed
    into thimble contributions
\begin{equation}\label{eq:thimbleDECOMPODITION}
\langle O \rangle = \frac{\sum_{\sigma} n_{\sigma} \,
  e^{-i\,S_I\left(p_{\sigma}\right)} \int_{\mathcal{J}_{\sigma}}
  \mathrm{d}z\, e^{-S_R} \,O\, e^{i\omega}}{\sum_{\sigma} n_{\sigma} \,
  e^{-i\,S_I\left(p_{\sigma}\right)} \int_{\mathcal{J}_{\sigma}}
  \mathrm{d}z\, e^{-S_R}  \, e^{i\omega}}
\end{equation}
the sum formally 
extending to all thimbles, even though some $n_{\sigma}$ can be zero
(thus, not all the critical points do contribute).
\end{itemize}
We still have to define what a thimble is, but we can immediately read
interesting features. A positive measure $e^{-S_R}$ is in place and constant
phases $e^{-i\,S_I\left(p_{\sigma}\right)}$ have been factored out of
the integrals: the imaginary part of the action stays
constant on each thimble. 
A so-called {\em residual phase} $e^{i \omega}$ is there: it
accounts for the relative orientation
between the canonical complex volume form and the real
volume form, characterizing the tangent space of the
thimble. \\
So, what is a thimble? It is the union of all {\em steepest ascent} (SA)
paths attached to a given critical point (we now explicitly write in
terms of the components of $z$)
$$
\frac{d}{dt}z_i = \frac{\partial \bar{S}}{\partial \bar{z}^i}
$$
It is easy to see that in this notation the critical point is
associated to $t=-\infty$. \\
A crucial point is that a thimble has the same real dimension of the
original domain of integration (on which the functional integral was
defined in the first instance). This can
be explicitly checked by solving the Takagi problem for the Hessian of
the action computed at the critical point
$$
H(S;p_\sigma)v_\sigma^{(i)}=\lambda_i^{(\sigma)}\bar{v}_\sigma^{(i)}
$$
{\em Takagi vectors} $v_\sigma^{(i)}$ provides a basis 
for the tangent space at the
critical point, while {\em Takagi values} $\lambda_i^{(\sigma)}$ fix 
the rate at which the
real part of action increases along the steepest ascent paths while
they leave the critical point. This is not the end of the story, since
we have only found out the tangent space at the critical point. In
order to get a basis at a generic point on the thimble we have to 
parallel-transport the basis for
the tangent space at the critical point along the flow.\\
Now, any flow (\ie any SA path) leaves the critical point 
along one possible direction on the tangent space. 
If we impose a normalization condition
$\sum_{i=1}^nn_i^2=\mathcal{R}$
all those directions are mapped to vectors 
$\sum_{i=1}^nn_iv^{(i)}$.
It is thus
quite natural to single out any given point on a thimble by the
correspondence
\begin{equation}
\label{eq:nNt}
\mathcal{J}_\sigma\ni z\leftrightarrow \left(\hat{n},t\right)\in S^{n-1}_{\mathcal{R}}\times\mathbb{R}
\end{equation}
with $S^{n-1}_{\mathcal{R}}$ the $(n-1)$-sphere of radius
$\sqrt{\mathcal{R}}$ and where we denote by $t$ the {\em time} coordinate 
parametrizing the flow along the SA path. 
In order to have a basis $V_{\sigma}^{(i)}(\hat{n},t)$ for the tangent
space at the (generic) point associated
to direction $\hat{n}$ and flow time $t$
one has to solve the (associated) flow equations
\begin{equation}
\label{eq:flowVECs}
\frac{\dd V_\sigma^{(j)}}{\dd
  t}=\sum_{i=1}^n\bar{V}_\sigma^{(i)}\,\overline{\l(\frac{\de^2S}{\de
    z^i\de z^j}\r)}.
\end{equation}

\subsection{A crude Monte Carlo on thimbles}\label{sec:thimbles_2}

By changing variables of integration one can now rephrase the thimble
decomposition as
\begin{equation}
\label{eq:Integrals_nt}
\langle O \rangle = \frac{\sum_{\sigma} n_{\sigma} \,
  e^{-i\,S_I\left(p_{\sigma}\right)} \int_{\sigma}\mathcal{D}\hat{n}\; 
2\sum_{i=1}^n\lambda_i^{(\sigma)}n_i^2
\int\limits_{-\infty}^{+\infty}\mathrm{d}
t\,e^{-S_{\mathrm{eff}}^{(\sigma)}(\hat{n},t)} 
\,O (\hat{n},t)\, e^{i\omega (\hat{n},t)}}{\sum_{\sigma} n_{\sigma} \,
  e^{-i\,S_I\left(p_{\sigma}\right)} \int_{\sigma}\mathcal{D}\hat{n}\; 
2\sum_{i=1}^n\lambda_i^{(\sigma)}n_i^2
\int\limits_{-\infty}^{+\infty}\mathrm{d}
t\,e^{-S_{\mathrm{eff}}^{(\sigma)}(\hat{n},t)} \, e^{i\omega
  (\hat{n},t)}}
\end{equation}
where we have defined
$$
S_{\mathrm{eff}}^{(\sigma)}(\hat{n},t)=S_R(\hat{n},t)-\log\left|\det V_{\sigma}(\hat{n},t)\right|\label{eq:Zn_Seff}.
$$
The formula has to be understood in the following way. 
The ${\lambda_i^{(\sigma)}>0}$ are the Takagi values (solutions
of the Takagi problem at the critical point $p_\sigma$) and 
at each critical point one has to solve a different
Takagi problem, resulting in different Takagi values
$\lambda_i^{(\sigma)}$ and different Takagi vectors $v_\sigma^{(i)}$, which are the
initial values for different $V_{\sigma}^{(i)}(\hat{n},t)$, solutions
of the flow equations (\ref{eq:flowVECs}).
One then assembles the $V_{\sigma}^{(i)}$ into the matrix $V_{\sigma}$,
the modulus of whose determinant enters the definition of 
$S_{\mathrm{eff}}$. At the same time, the phase of 
$\det V_{\sigma}(\hat{n},t)$ provides
the residual phase $e^{i\omega(\hat{n},t)}$.
The notation $\int_{\sigma}$ shows that 
(\ref{eq:nNt}) holds {\em for each critical point} and 
at each critical point one has to solve a different
Takagi problem, resulting in different
$\Delta_{\hat{n}}^{(\sigma)}(t)$.\\

\noindent
We can now devise a simple crude Monte Carlo scheme for simulating on
thimbles:
\begin{itemize}
\item We pick up randomly (with flat distribution) a direction $\hat{n}$. 
\item Since we want to compute the contribution coming from the SA leaving
  the critical point $p_\sigma$ along $\hat{n}$, we 
  prepare convenient {\em initial conditions} both for the field and
  for the tangent space basis vectors for such a SA. We can do this, 
since near the critical point
  solutions of the flow equations are know as\footnote{For details see
    \eg \cite{DiRenzo:2015foa}.}
\begin{eqnarray}
z_j\left(t\right) & \approx & z_{\sigma,j}+\sum\limits_{i=1}^n
                              n_i  \; v_{\sigma j}^{\left(i\right)} \;
                              e^{\lambda_i^{(\sigma)} t} \nonumber \\
 V_{\sigma j}^{(i)}\left(t\right) & \approx &
v_{\sigma j}^{\left(i\right)} \; e^{\lambda_i^{(\sigma)} t} \nonumber
\end{eqnarray}
which we can compute for $t=t_0\ll0$.
\item We then integrate the SA equations for the field and 
the equations for transporting the basis vectors all the way up till 
we reconstruct the ($\mathrm{d}t$) 
integrals appearing in (\ref{eq:Integrals_nt}) and while 
ascending we compute both the integral in the numerator and the
one in the denominator.
\end{itemize}

\section{Thimble regularization for $0+1$ dimensional QCD}\label{sec:QCD01}

\subsection{QCD in $0+1$ dimensions}\label{QCD01}

QCD in $0+1$ dimensions is given in terms of
\emph{staggered} fermions on a one-dimensional lattice 
with (even) $N_t$ sites in the temporal direction (the 
temperature being given by $aN_t=1/T$, where $a$ is the lattice 
spacing). 
A genuine sign problem is there as in real 
QCD, due to the presence of a (quark) chemical potential. 
The partition function of the theory for $N_f$ degenerate 
quark flavours of mass $m$ is
\begin{equation*}
Z_{N_f}=\int\prod\limits_{i=1}^{N_t}\dd U_i\,{\det}^{N_f}(aD)
\end{equation*}
where $D$ is the lattice staggered Dirac operator
\begin{equation*}
(aD)_{ii'}=am\delta_{ii'}+\frac{1}{2}\l(e^{a\mu}U_i\tilde{\delta}_{i',i+1}-e^{-a\mu}U^\dag_{i-1}\tilde{\delta}_{i',i-1}\r)
\end{equation*}
and $\tilde{\delta}_{ii'}$ is the anti-periodic Kronecker delta. 
In a convenient gauge one actually has
\begin{equation*}
Z_{N_f}=\int\limits_{\mrm{SU}(3)}\dd U\,e^{-S(U)}
\end{equation*}
where
\begin{equation*}
S(U)=-N_f\tr\log M(U) = -N_f\tr\log \left(A\,\id_{3\times 3}+e^{\mu/T}U+e^{-\mu/T}U^{-1}\right)
\end{equation*}
while $A=2\cosh(\mu_c/T)$ and $\mu_c=\sinh^{-1}(m)$ (from now on, we 
set $a=1$). 
We computed three main observables. The chiral
condensate is the first one 
\begin{equation*}
\Sigma\equiv T\frac{\de}{\de m}\log Z=T\l\langle N_f\tr\l(M^{-1}\frac{\de M}{\de m}\r)\r\rangle=N_f\sqrt{\frac{A^2-4}{m^2+1}}\l\langle\tr\l(M^{-1}\r)\r\rangle.
\end{equation*}
The other two are the Polyakov loop $\langle\tr\,U\rangle$ and the
anti-Polyakov loop
$\langle\tr\,U^\dag\rangle=\langle\tr\,U\rangle_{\mu\rightarrow-\mu}$. The
latter two can be related to the quark number density $n\equiv
T\frac{\de}{\de\mu}\log Z$ by a relation which takes quite different
forms for different values of $N_f$. 
There are known numbers to compare to, since 
analytical results for $0+1$ QCD are available
(see for example \cite{Bloch:2013ara}).

\subsection{$0+1$ QCD in thimble regularization}\label{sec:thimbleQCD01}

\begin{figure}[htb] 
  \centering
  \subfloat[$m=0.1$, $N_f=1$]{
    \label{fig:qcd1_cc_m01_1}
    \includegraphics[height=4.5cm]{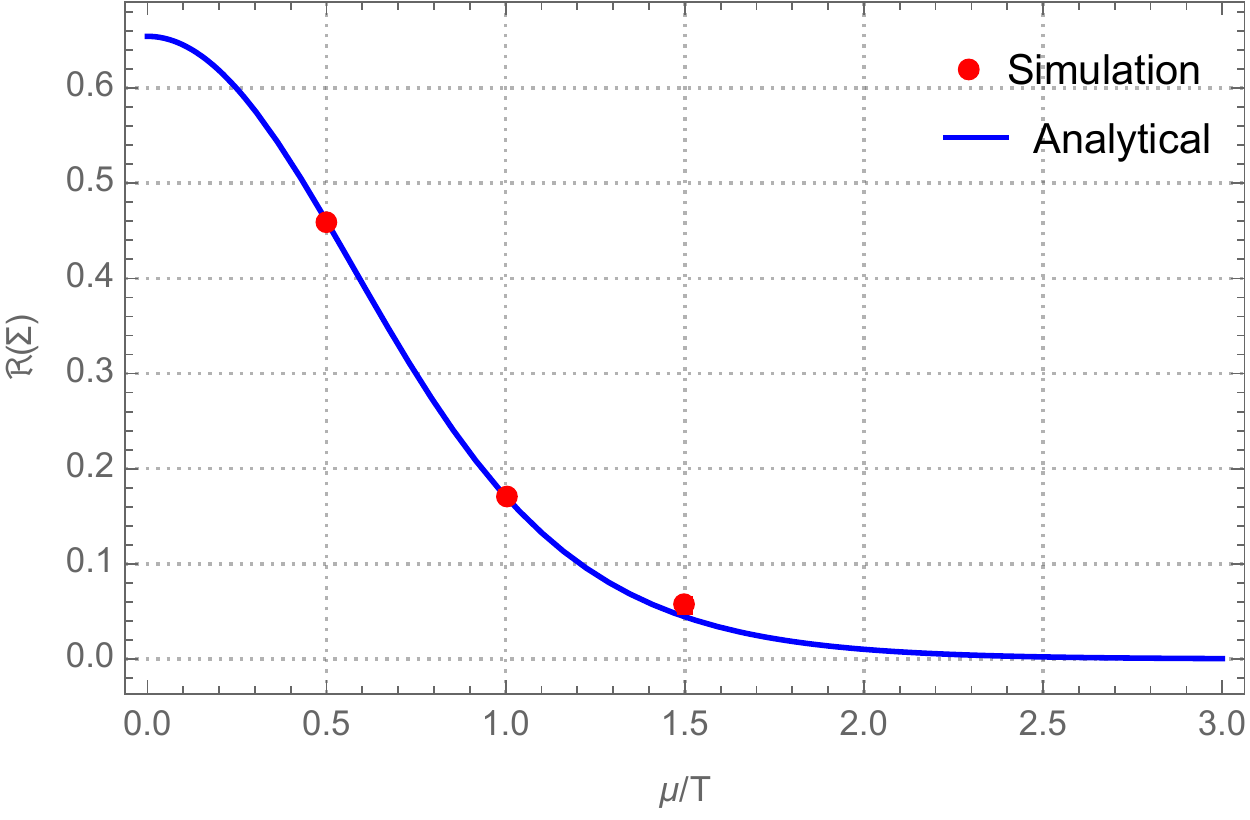}
  }
  \subfloat[$m=0.1$, $N_f=12$]{
    \label{fig:qcd1_cc_m01_12}
    \includegraphics[height=4.5cm]{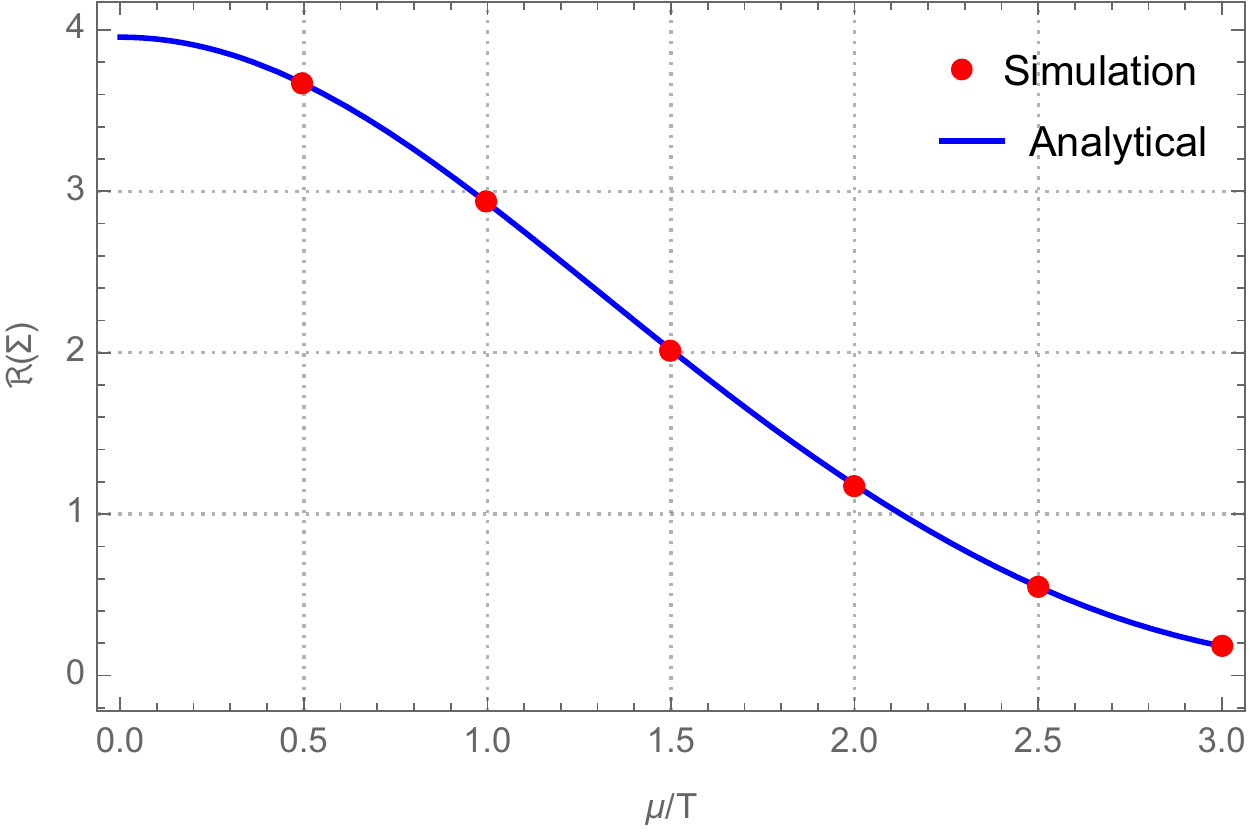}
  }
  \caption{Situation one year ago: (right) chiral condensate correctly
    computed for high $N_f$ (here $N_f=12$); (left) for $N_f=1$ 
we could not get results beyond a given value of $\mu/T$.}
  \label{fig:snort}
\end{figure}

To solve in thimble regularization we need two ingredients:
\begin{itemize}
\item We need to 
complexify the degrees of freedom, \ie
$$
\mrm{SU}\l(N\r)\ni U=e^{i x_aT^a}\rightarrow e^{i z_aT^a}=e^{i \l(x_a+i y_a\r)T^a}\in\mrm{SL}\l(N,\mbb{C}\r).
$$
Notice that 
$$
\mathrm{SU}\left(N\right)\ni U^\dag=e^{-i x_aT^a}\rightarrow e^{-i z_aT^a}=e^{-i \left(x_a+i y_a\right)T^a}=U^{-1}\in\mathrm{SL}\left(N,\mathbb{C}\right).
$$
\item We write the SA equations (in terms of Lie derivatives) as 
\beq \label{eq:SAgauge}
\frac{\mrm{d}}{\mrm{d}\tau}U_\muh\l(n;\tau\r)=\l(i\,T^a\bar{\nabla}_{n,\muh}^a\overline{S\l[U\l(\tau\r)\r]}\r)U_\muh\l(n;\tau\r).
\eeq
\end{itemize}
The equations for the Hessian and the flow equations for the vectors
can also be easily written down\footnote{For all the details see~\cite{DiRenzo:2017igr}
  which has been recently issued.}. 
Notice that we find three critical points $\{U_k=e^{2\pi i k/3}\id\}$ with
$k=0,1,2$ and they all contribute to the thimble decomposition.\\
At last year conference we presented results~\cite{DiRenzo:2016pwd}, 
which were obtained
using the crude Monte Carlo scheme presented above. We took into
account all the three critical point, even if one can notice that
there are regions of the parameter space where the thimble attached to the
identity captures virtually the complete result. This agrees with 
{\em semiclassical estimates}. The complete partition function is
built out of the three contributions attached to the critical points
$Z=Z_0+Z_1+Z_2$
and one can define and compute (in the semiclassical approximation)
$$
r^{1,2}_0\equiv\frac{\l|Z_{1,2}\r|}{\l|Z_0\r|}
$$
which tells us the relative weights of the Z's. \\
All in all, we were able to compute {\em almost} everything in a wide
region of the parameter space covering a range of values for
$N_f$, $m$ and $\mu/T$. {\em Almost} means that at high values of 
$N_f$, flat Monte Carlo simulations were successful at all values of
$\mu/T$ (this is consistent with the observation that the model is easy 
to simulate at high $N_f$; in those regions the problem is isotropic
and semiclassical estimates essentially become exact 
in the limit $N_f\rightarrow\infty$). On the
other side, for other values of parameters (namely, large $\mu/T$ at 
small $N_f$) results could not be calculated (see
Figure~\ref{fig:snort}). 

\section{Improving on previous results}\label{sec:DObetter}

\subsection{Reflection symmetry}\label{sec:symmetry}

Simmetries always play an important role: not surprisingly, we have a
symmetry at hand which lets us check our results and which also makes
our life easier~\cite{Tanizaki:2015pua}.
The QCD action (in our case, we only have the Dirac determinant) 
is invariant under charge conjugation $\mcal{C}$ defined by 
\begin{align*}
\mcal{C}
\begin{cases}
\psi\rightarrow C^{-1}\bar{\psi}^T\\
\bar{\psi}\rightarrow-\psi^T C\\
U_\nuh(n)\rightarrow\bar{U}_\nuh(n) & \l(A_\nuh(n)\rightarrow-A^T_\nuh(n)=-\bar{A}\r)\\
\mu\rightarrow-\mu
\end{cases}
\end{align*}
Together with the generalization of $\gamma_5$-hermiticity at finite chemical potential 
\begin{equation*}
\det D(U,-\mu)=\overline{\det D(U,\mu)}
\end{equation*}
this implies that
\begin{equation*}
\overline{S(A)}\sim\overline{\det D(U,\mu)}\overset{\gamma_5\text{-herm.}}{=}\det D(U,-\mu)\overset{\mcal{C}\text{-inv.}}{=}\det D(\bar{U},\mu)\sim S(-\bar{A})
\end{equation*}
We thus expect thimbles to appear in conjugate pairs. Consider the 
$\{U_k\}$. $U_0=\id$ is real and therefore self-conjugate (\ie 
computations on the associated thimble yield real results). 
Being
$e^{4\pi i/3}=e^{-2\pi i/3}$, we immediately see that
$U_2=\overline{U_1}$: $U_1$ and $U_2$ are a
conjugate pair of critical points and results of integration on $U_2$
should be the complex conjugate of those on $U_1$, yielding an overall
real contribution to the partition function and to the
expectation value of observables as well. 
All this is well evident in numerical simulations and in all 
the numerical results that we present in the following we take 
advantage of this symmetry: the real part of results from thimble 
1 and 2 have been averaged.

\subsection{Importance sampling}\label{sec:ImpSampl}

If a single thimble contributed\footnote{For the sake of
notational simplicity we will often omit in the following the
subscript/superscript $\sigma$, \eg in Takagi values.}, 
the computation of (\ref{eq:thimbleDECOMPODITION}) would simply amount 
to\footnote{Notice that reweighting with respect to the critical phase is
  in place.}
\begin{equation}
\langle O\rangle=\frac{\langle O\,e^{i\,\omega}\rangle_\sigma}{\langle
  e^{i\,\omega}\rangle_\sigma}
\;\;\;\;\;\;\;\;\; \mbox{with} \;\;\;\;\;\;\;\;\;
\langle\ldots\rangle_\sigma=\frac{\int\limits_{\mathcal{J}_\sigma}\mathrm{d}^ny\,\ldots\,e^{-S_R}}
{\int\limits_{\mathcal{J}_\sigma}\mathrm{d}^ny\,e^{-S_R}}
\label{eq:obs_reweighted}
\end{equation}
In terms of the representation (\ref{eq:nNt}), we can now rephrase
\begin{equation}
\langle f \rangle_\sigma=
\frac{1}{Z_\sigma}\int\limits_{\mathcal{J}_\sigma}\mathrm{d}^ny\,f\,e^{-S_R}=\frac{1}{Z_\sigma}\int\mathrm{D}\hat{n}\;
(2\sum_{i=1}^n\lambda_in_i^2) \int\limits_{-\infty}^{+\infty}
\mathrm{d}t\,f(\hat{n},t)\,
e^{-S_{\mathrm{eff}}(\hat{n},t)}=\int\mathrm{D}\hat{n}\,\frac{Z_{\hat{n}}}{Z_\sigma}\,f_{\hat{n}}\label{eq:MC_expect_f}
\end{equation}
where
\[
f_{\hat{n}}\equiv\frac{1}{Z_{\hat{n}}}
(2\sum_{i=1}^n\lambda_in_i^2) \int\limits_{-\infty}^{+\infty}
\mathrm{d}t\,f(\hat{n},t)\,
e^{-S_{\mathrm{eff}}(\hat{n},t)}
\;\;\;\;\;\;\;\;\; \mbox{and} \;\;\;\;\;\;\;\;\;
Z_{\hat{n}}^{(\sigma)}=2\sum_{i=1}^n\lambda_i^{(\sigma)}n_i^2\int\limits_{-\infty}^{+\infty}\mathrm{d} t\,e^{-S_{\mathrm{eff}}^{(\sigma)}(\hat{n},t)}.
\]
Now, $f_{\hat{n}}$ looks like a functional integral along a single complete flow
line. On the other side, 
(\ref{eq:MC_expect_f}) is nothing but the average of the
$f_{\hat{n}}$ ({\em i.e.} the average of the contributions that a 
given observable takes from complete flow lines) and the weight 
$Z_{\hat{n}}/Z_\sigma$ represents the fraction of the
partition function which is provided by a single complete flow line.
$Z_{\hat{n}}/Z_\sigma$ provides a natural setting
for importance sampling: {\em directions} $\hat{n}$ have to be
extracted according to the probability 
$P(\hat{n})=Z_{\hat{n}}/Z_\sigma$. 

We thus proceed as follow. 
Sitting on the current configuration 
(associated to a direction $\hat{n}$), we propose a new one 
(associated to a direction $\hat{n}'$). $\hat{n}'$ is identical to $\hat{n}$ 
apart from two randomly chosen components, 
say $(n_i,n_j)$ with $i\neq j$. 
Given the normalization
$\left|\vec{n}\right|=\sqrt{\mathcal{R}}$ and the values of all
$\{n_k\}_{k\neq i,j}$ we define $C$ as 
\[
C\equiv n_i^2+n_j^2=\mathcal{R}-\sum_{k\neq i,j}n_k^2
\]
and we observe that there is a coordinate system in which  
the current values of $(n_i,n_j)$ are parametrized as
\[
n_i=\sqrt{C}\,\cos\phi \;\;\;\;\; n_j=\sqrt{C}\,\sin\phi
\]
with $\phi\in[0,2\pi)$. Now we change 
$\phi \rightarrow \phi'$ by extracting $\phi'-\phi$ flat 
in a given (tunable) range.
This results in $(n_i,n_j) \rightarrow 
(n_i',n_j')$, while for all the other components (${k\neq i,j}$) 
$n_k = n_k'$. 
We finally accept the proposed configuration with the
standard Metropolis test
\begin{equation}
P_{\mathrm{acc}}\left(\hat{n}'\bigr|\hat{n}\right)=
\min\left\{1,\frac{Z_{\hat{n}'}}{Z_{\hat{n}}}
\right\}.\label{eq:MC_flatmetroacc}
\end{equation}\\

\begin{figure}[ht] 
  \centering
  \subfloat[$m=0.1$, $N_f=1$]{
    \label{fig:qcd1_cc_m01_1}
    \includegraphics[height=4.5cm]{qcd1_cc_m01_1.pdf}
  }
  \subfloat[$m=0.1$, $N_f=1$ (new!)]{
    \label{fig:qcd1_cc_m01_1_new}
    \includegraphics[height=4.5cm]{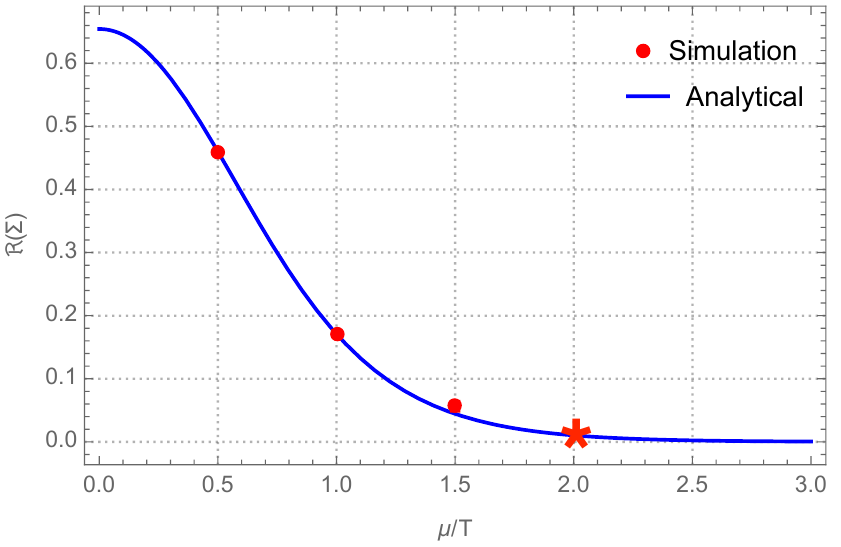}
  }
  \caption{An example of succesfull improvement: (left) as in Figure \ref{fig:snort}; 
(right) a new point computed via the improved method.}
  \label{fig:yap}
\end{figure}

In our case three contributions should be in principle taken into
account. Actually, due to the symmetry of Section \ref{sec:symmetry},
only two distinct contributions are in place and
we have to compute
\begin{equation}
\langle O\rangle=\frac
{n_0 \, e^{-i\,S_{I 0}} \, Z_0 \, \langle O\,e^{i\,\omega}\rangle_0 \, + \, n_{12}
  \, e^{-i\,S_{I 12}} \, Z_{12} \, \langle O\,e^{i\,\omega}\rangle_{12}}
{n_0 \, e^{-i\,S_{I 0}} \, Z_0 \, \langle e^{i\,\omega}\rangle_0 \, + \, n_{12}
  \, e^{-i\,S_{I 12}} \, Z_{12} \, \langle e^{i\,\omega}\rangle_{12}}
\label{eq:obs_reweighted_2}
\end{equation}
in which subscript notations should be evident. In~\cite{DiRenzo:2015foa}
we observed that equation (\ref{eq:obs_reweighted_2}) can be rewritten 
\begin{equation}
\langle O\rangle=\frac
{\langle O\,e^{i\,\omega}\rangle_0 \, + \, \alpha
\langle O\,e^{i\,\omega}\rangle_{12}}
{\langle e^{i\,\omega}\rangle_0 \, + \, \alpha
\langle e^{i\,\omega}\rangle_{12}}
\label{eq:obs_reweighted_3}
\end{equation}
where we defined 
\begin{equation}
\alpha \equiv \frac
{n_{12} \, e^{-i\,S_{I 12}} \, Z_{12}}
{n_0 \, e^{-i\,S_{I 0}} \, Z_0}
\label{eq:DEFalpha}
\end{equation}
The idea is now to take a given
observable as a {\em normalization point}, thus 
determining the value of $\alpha$. All the other observables of
the theory can then be computed using this input. This could indeed be
successfully done. 
Figure \ref{fig:yap} confirms the effectiveness of the
procedure. For $N_f=1, m=0.1$ the point $\mu/T=2.0$ was completely out
of reach for flat, crude Monte Carlo, while results are successfully
computed with the improved method. At these
values of parameters the tiny value of the chiral condensate (of order
$10^{-2}$) results from a delicate cancelation of the contributions 
coming from the different thimbles. This has nothing to do with the 
original sign problem (nor \eg with the residual phase), but it is 
simply a numerical 
accident occurring for a given observable at a given value of
parameters. 

\section{Conclusions}
There is still quite a long way to go before thimble regularization
can tackle the real goal, \ie QCD at finite density. This work
nevertheless shows that simulations on multiple thimbles are viable.

\section*{Acknowledgments}
It is a pleasure for the author to thank Giovanni Eruzzi, in collaboration with whom
these results have been worked out. The author acknowledges support by
I.N.F.N. under {\em i.s.} QCDLAT.

\bibliography{lattice2017fdr}

\end{document}